# An Optimal Piezoelectric Beam for Acoustic Energy Harvesting

Amir Panahi [1,*], Alireza Hassanzadeh [1], Ali Moulavi [1], and Ata Golparvar [2]

[1] Faculty of Electrical Engineering, Shahid Beheshti University, Tehran, Iran.

[2] Faculty of Engineering and Natural Sciences, Sabanci University, Istanbul, Turkey.

**Correspondence**

Amir Panahi, Faculty of Electrical Engineering, Shahid Beheshti University, Tehran, Iran.

Email: am.panahi@mail.sbu.ac.ir

**Summary:** This study presents a novel piezoelectric beam structure for acoustic energy harvesting. The beams have been designed to maximize output energy in areas where the noise level is loud such as highway traffic. The beam consists of two layers (copper and polyvinylidene fluoride) that convert the ambient noise's vibration energy to electrical energy. The piezoelectric material's optimum placement have been studied, and its best positon is obtained on the substrate for the maximum yield. Unlike previous studies, which the entire beam substrate used to be covered by a material, this study presents a modest material usage and contributes to lowering the harvester's final production cost. Additionally, in this study, an electrical model was developed for the sensor and a read-out circuitry was proposed for the converter. Moreover, the sensor was validated at different noise levels at various lengths and locations. The simulations were performed in COMSOL Multiphysics® and MATLAB® and report a maximum sound pressure of 140 dB from 100 dB point sources in an enclosed air-filled cubic meter chamber.

## 1. INTRODUCTION

The increase of energy consumption in portable electronics ranging from smartphones to internet-of-thing (IoT) nodes and wearable devices are all urgently requiring new energy resources and storage systems so that the trend in integration and miniaturization could be continued.[1-5] One of the known energy harvesting methods is exploiting the phenomena of molecule polarization by electric surface charges and generation of piezoelectric voltage [6]. Subsequently, piezoelectric materials can be triggered by noise or mechanical forces, and attracted much attention due to their natural characteristics of compactness, simple modeling, scalability, and compatibility with integrated circuits [7]. Harvested energy of piezoelectric materials can be used to power wireless sensor networks [8]. On the other hand, power transmission from a power plant and storing it in batteries may be very costly due to limited access to some locations or unfavorable weather conditions. Finding a solution to this issue has been a significant challenge for researchers and environmental energy resources, such as sunlight, wind, noise, and motion (i.e., displacement), was considered [9]. Ambient noise is heard every day in metropolitan areas, and they are among the first to exploit it in energy harvesting applications. For instance, zinc oxide (ZnO) nanowires were used to convert vibrational energy into electrical power, or similarly, layers of brass, aluminum, and lead zirconate titanate (PZT) were shown as promising materials for energy harvesting [10,11]. Additionally, to increase efficacy, a proof mass at the beam's tip was proposed to reduce the resonance frequency [12].

Moreover, the genetic algorithm was previously acquired to optimize the mass's location on the beam to obtain the highest power levels [14]. Instead, this study pro-poses an optimized energy harvester beam triggered by ambient noise using copper and piezoelectric polyvinylidene fluoride (PVDF) layers. Here, beam lengths and resonance frequencies are customized. The beams' length is directly related to the vibrational resonance frequency, so optimizing the beam length in line with the resonant frequency can lead to more voltage harvesting. Additionally, the piezoelectric layer's best position is investigated to the function applied strain by the copper layer to the piezoelectric material so that the material cost could be further reduced.

## 2. VIBRATION ANALYSIS AND FUNDAMENTAL FREQUENCY ESTIMATION

Principles of vibration and sound comprise studying the oscillatory behavior of objects. In short, any elastic solid can vibrate and resist an applied force, heat, and deformation and return to its initial state if the acting factors are removed. In this section, the equation of motion is derived using the Euler-Bernoulli beam theory.

The motion equation for a uniform beam under free vibration is shown in Equation (1) [15, 16].

$$EI\frac{\partial^4 z(x.t)}{\partial x^4} + m\frac{\partial^2 z(x.t)}{\partial t^2} = 0 \tag{1}$$

Where $EI$ stands for the bending strength, $m$ is the beam's mass, and $z(x,t)$ is of the transverse displacement of the neutral axis due to bending and expressed in Equation (2).

$$z(x.t) = z_b(x.t) = z_{rel}(x.t) \tag{2}$$

Also, $z_b(x,t)$ is the displacement of the beam base, $z_{rel}(x,t)$ is the transverse displacement of the beam relative to the base. In free vibration, the transverse displacement of rectangular beams is given in Equation (3):

$$z_{rel}(x.t) = \sum_{k=1}^{\infty} \omega_k(x) q_k(t) \tag{3}$$

Where $\omega_k(x)$ and $q_k(t)$ are the optimal normalized mass and coordinate the function of a free and single-ended beam in the $k$ (i.e., mode shape parameter), respectively. With substituting Equation (3) to Equation (1), Equation (4) is archived.

$$\frac{EI}{m\omega_k(x)}\frac{d^4 \omega_k(x)}{dx^4} = -\frac{1}{q_k(t)}\frac{d^2 q_k(t)}{dt^2} = \omega_k^2 \tag{4}$$

Where $\omega_k^2$ is a positive constant in response to harmonic. The left side equation in (4) is reduced to Equation (5).

$$\frac{d^4 \omega_k(x)}{dx^4} - \lambda_k^4 w_k(x) = 0 \tag{5}$$

Where $\lambda_k$ is defined in Equation (6):

$$\lambda_k^4 = \frac{m}{EI}\omega_k^2 \tag{6}$$

By assuming $\omega_k(x) = Ce^{sx}$ (where $s$ and $C$ are constant coefficients), Equation (5) could be solved by Equation (7).

$$\omega_k(x) = C_1 \sin(\frac{\lambda_k}{L}x) + C_2 \cos(\frac{\lambda_k}{L}x) + C_3 \sinh(\frac{\lambda_k}{L}x) + C_4 \cosh(\frac{\lambda_k}{L}x)$$

Where constant coefficients $C1, C2, C3$, and $C4$ are obtained from boundary conditions and $\omega_k$ denotes the beam's natural frequency in $k$ mode in Equation (8).

$$\omega_k = \lambda_k^2 \sqrt{\frac{EI}{mL^4}} \tag{8}$$

Since two different materials are used for constructing this beam, $EI$ and $m$ computed in Equation (9) [5]:

$$m = B(\rho_b h_b + \rho_p h_p) \tag{9}$$

$$I = B\left[\begin{array}{l} E_b\left(\dfrac{h_b^3}{12}\right) \\ +\dfrac{E_p}{3}\left(\left(\dfrac{h_b}{2}+h_p\right)^3 - \dfrac{h_b^3}{8}\right) \end{array}\right] \tag{10}$$

Where $b$, $\rho$ and $h$ denote the width, density, and thickness, respectively. Using modal analysis and boundary conditions in Equation (7), the frequency equation is obtained Equation (11), and considering Equation (7), Equation (12) is achieved.

$$1 + \cos \lambda_k \cosh \lambda_k = 0 \tag{11}$$

$$\omega_k(x) = \sin\left(\frac{\lambda_k}{L}x\right) - \sinh\left(\frac{\lambda_k}{L}x\right) + \beta_k\left[\cos\left(\frac{\lambda_k}{L}x\right) - \cosh\left(\frac{\lambda_k}{L}x\right)\right] \tag{12}$$

Where $\beta_k$ is computed in Equation (13).

$$\beta_k = -\frac{\cos \lambda_k + \cosh \lambda_k}{\sin \lambda_k - \sinh \lambda_k} \tag{13}$$

The natural frequency is the frequency at which the system predominantly tends to oscillate. The optimal fundamental frequency yields the highest voltage and power in rectangular beams.

## 3. PROPOSED PIEZOELECTRIC BEAM

The beam structure is mainly used for piezoelectric energy harvesting that deflects under vibration loads. The rectangular beam has been shown in Figure 1 that is designed and simulated in COMSOL Multiphysics in finite element method (FEM) with a fixed end. Usage of several layers will reduce the resonance frequency in piezoelectric materials [17]. Figure 1 illustrates a double-layer structure where its top layer is polyvinylidene fluoride (PVDF) and the substrate is copper. The structure is single-ended on its left, and the beam is free to vibrate on the right side; thus, the vibration excites PVDF material at the end of the copper beam. The piezoelectric layer is located at the optimal point on the copper layer, i.e., where most of the strain is applied to the PVDF material. The PVDF polymer is used instead of PZT due to its higher sensitivity [17, 18]. Therefore, multi-layer structures are recommended to reduce oscillation frequency.

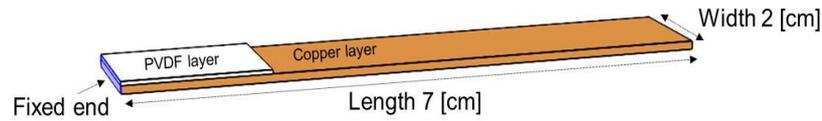

Figure 1. Proposed piezoelectric cantilever fixed end; PVDF and copper layers are colored with white and brown, respectfully; The fixed end clamp is colored blue.

For the proposed beam, the length, width, and thickness of the two layers and the system settings are shown in Table 1. Proposed dimensions were implemented, and vibration mode shapes are calculated to obtain the beam's natural frequency and presented in Table 2. Here, horizontal, vertical, and edge vibrations are investigated. The frequency of vertical vibration is 121.91 Hz is the most practical natural frequency for this system; it causes higher voltage and more charge on the surface of the piezoelectric beam [13]. The two layers act as an integrated (i.e., unified) unit and vibrate at the vibration simulation's natural frequency.

## 4. SIMULATION & THE READ-OUT CIRCUITRY

Simulation results of the beam structure with the FEM model in this section are shown in the tables and figures. Additionally, an interface circuit model for harvesting energy of the beam is presented. Material properties and detail of each layer as length, width, thickness, young's modulus and density in the Table 1 is presented.

**Table 1.** Material properties and dimensions of layers used in the rectangular beam.

| Copper beam parameters | Value | PVDF beam parameters | Value |
|---|---|---|---|
| Length [m] | 0.07 | Length [m] | 0.02 |
| Width [m] | 0.02 | Width [m] | 0.02 |
| Thickness [m] | 0.001 | Thickness [m] | 0.0005 |
| Young's modulus [N/m2] | $110*10^9$ | Young's modulus [N/m2] | $1.2*10^9$ |
| Density (ρ) [Kg/m3] | 8960 | (Density (ρ) [Kg/m3] | 1780 |

**Table 2.** Natural frequencies of the beam with different vibration modes.

| Vibration mode | Mode 1 | Mode 2 | Mode 3 | Mode 4 | Mode 5 | Mode 6 |
|---|---|---|---|---|---|---|
| Natural frequencies [Hz] | 121.91 | 745.8 | 827.61 | 2074.3 | 2193.8 | 2551.7 |

Six different resonance modes for a 7 cm piezoelectric beam is presented in Table 2.

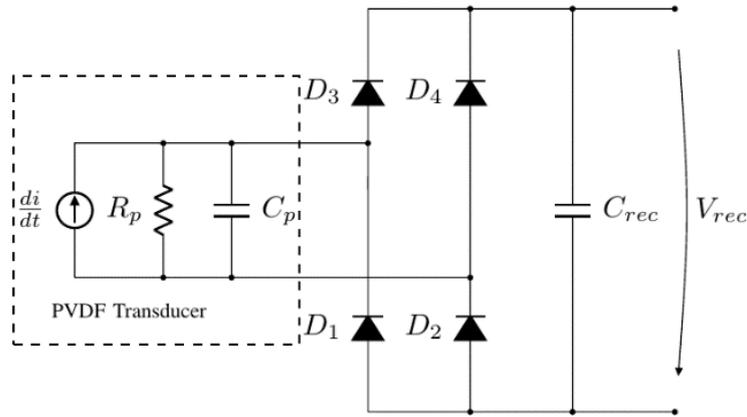

Figure 2. A typical piezoelectric transducer and AC-DC converter.

Figure 2 illustrates the piezoelectric transducer circuit model. The circuit includes a piezoelectric transformer circuit, a diode bridge rectifier, and a storage capacitor. The leakage resistance is $R_p$ that because of dimension of PVDF has considerable value. $C_p$ is equal to 0.53 nF that is the PVDF layer's capacitance.

To determine the maximum effect of vibration on the beam, the optimal frequency at which the maximum displacement and excitation of the beam (i.e., more power generation) were achieved and illustrated in Figure 3. The beam (red section) has the highest massive vertical displacement, which creates the maximum strain over the overlap with the piezoelectric beam. The piezoelectric beam design is investigated by considering the beam length and its response to the generated noise.

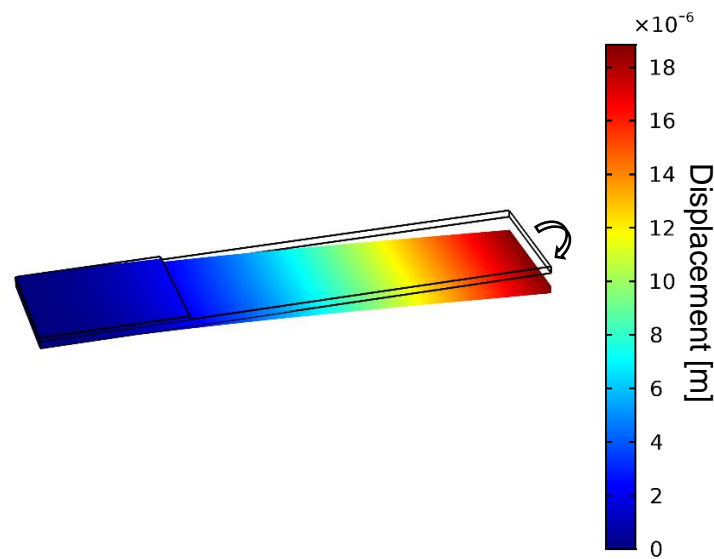

Figure 3. Displacement of the beam in the first resonance frequency.

## 6. DISCUSSION

By tuning the copper beam's length, the system is optimized to the highest harmonic frequency using the beam-sized simulation results at different recorded harmonic frequency amplitudes. A beam can be dedicated to each noise source.

In this work, to acquire and record the noises, a recorder (SONY ICD-UX560) for experimenting analysis is used for recording in four selected environments. These data were saved at a record mode of LPCM 44.1 kHz/16bit, which all of the data were saved. Figure 4 shows the power amplitude of experimentally recorded noises in terms of frequency. Figure 5 illustrates that, by decreasing the beam length, the output voltage decreases while the beam resonance frequency increases.

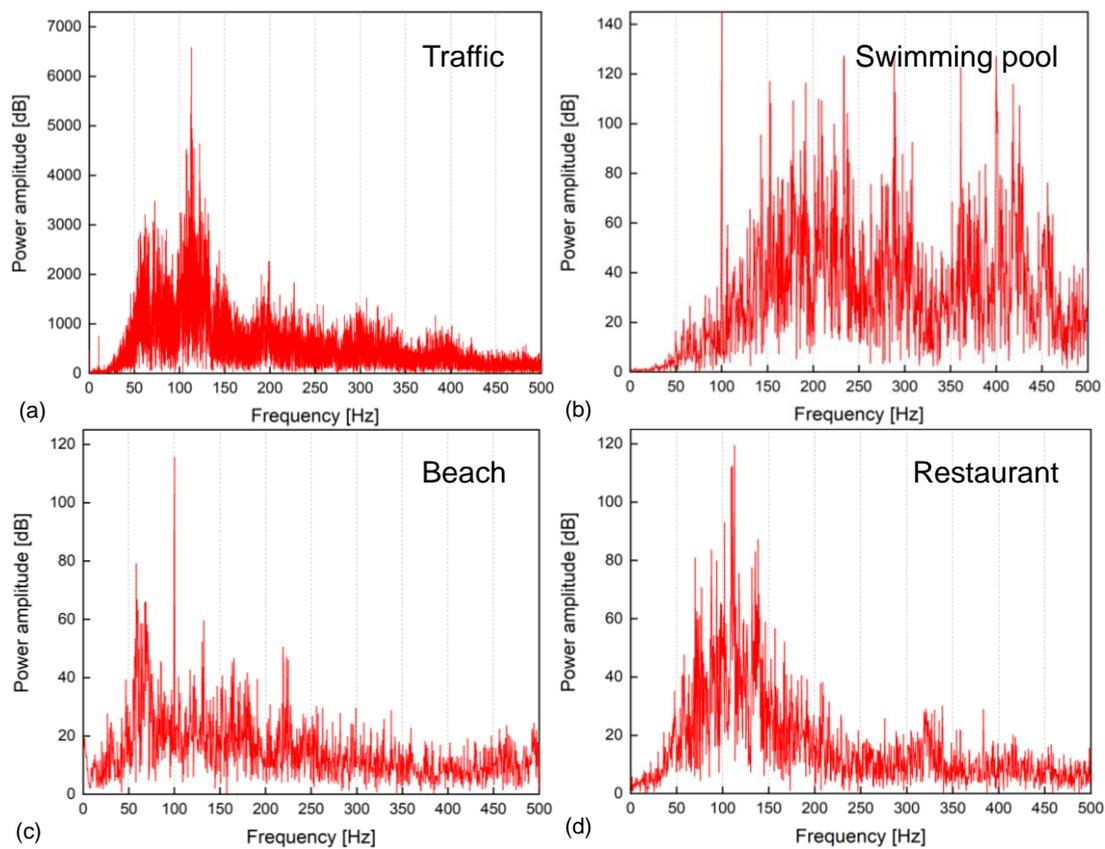

Figure 4. illustrates the fast Fourier transform of four sound signals from different locations, and the highest harmonic frequency increases from a crowded restaurant to a massive traffic street, respectively.

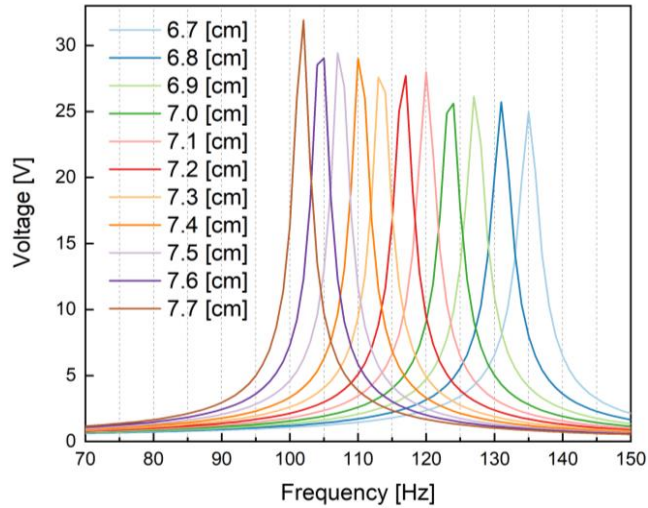

Figure 5. illustrates that the first harmonic frequency increases from a crowded restaurant to a traffic street. For instance, the most significant harmonic frequencies in a crowded restaurant, beach, swimming pool, and heavy-traffic street reach 105 Hz, 108 Hz, 114 Hz, and 120 Hz, respectively.

As the copper beam length increases, the resonance frequency of the beam decreases. Consequently, to harvest the maximum energy from noises with various frequencies, multiple beams of different lengths have to be connected to the same base; furthermore, a dedicated interface circuit was designed for each energy harvesting system. The thickness, width, and amount of piezoelectric material of all four beams are kept the same, but copper beam length differs from a location to another. In the modeling near the first fundamental frequency (121 kHz), a beam is placed at the center of an enclosed air-filled one cubic meter chamber that contains two sound sources with a power of 100 dB at the lower corners of the chamber with a sound speed of 343 m/s. The sound pressure distribution and acoustic streaming field in the modeled environment are presented in Figure 6.

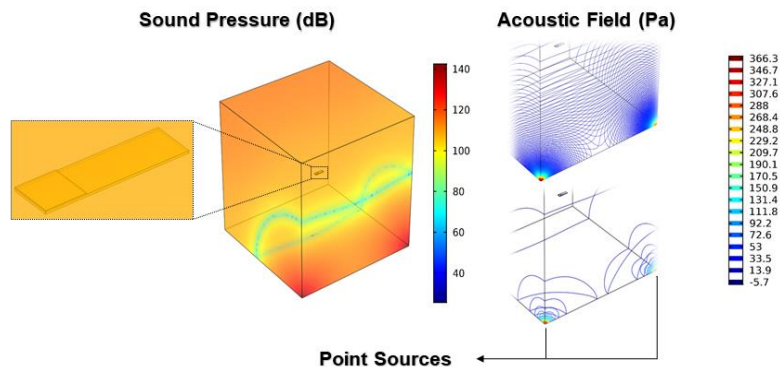

Figure 6. The sound pressure level distribution and acoustic field with two sources points.

## 7. CONCLUSION

In this paper, sound signals from four different locations (a crowded restaurant, beach, swimming pool, and heavy traffic street) recorded and were fast fourier transformed (FFT). From the FFT data of each of the four signals, a frequency with the highest weight was selected as the resonance frequency. At each position, a beam of beams of different lengths was simulated in COMSOL Multiphysics®. The length of the beam affects the first mode shape. Moreover, in this paper, the maximum voltage is obtained at the output by equating the beam's fundamental frequency with the sound resonance frequency. For a 7 cm beam, the sound pressure distribution was shown, an interface circuit was used to store the generated power so that a diode bridge rectifier rectifies the oscillating voltage generated by the piezoelectric material. Then, a parallel capacitor compensates the ripple voltage and provides a constant DC voltage for battery charging.